# A Gauss-Vaníček Spectral Analysis of the Sepkoski Compendium: No New Life Cycles

*New periods can emerge from data as a byproduct of incorrect processing or even the method applied. A good way to avoid this error is to use Gauss-Vaníček spectral analysis. It easily detects periods in raw and gapped records, and in physical sciences, it can detect eigenfrequencies and relative dynamics accurately and simultaneously.*

More often than they should, reports in the open literature on data processing involve inappropriate numerical techniques for the data of interest. Robert A. Rohde and Richard A. Muller,[1] for example, recently claimed a new 62 ± 3 million years (Myr) period from a Fourier spectral analysis (FSA) of genera variations "about" a (third-order polynomial) trend; essentially, the authors modified genera variations from the *Sepkoski compendium*, the world's most complete fossil record.[2] Rohde and Muller interpreted their alleged finding not as periodicity about a subjectively presumed trend—justifiable only within a specific technique's narrow application domain—but as the periodicity of life itself. Thus, they uncritically attached universality to the realm of an imperfect numerical analysis technique.

The purpose of this article is to verify Rohde and Muller's results by analyzing the exact same sequence (provided by Rohde) of well-resolved genera as the time series used to produce Figure 1 (which is a reprint of Figure 1e from their original article).[1] Admittedly, the technique described here and the technique that Rohde and Muller used might not belong to the same general application domains, but that's precisely what makes this study an authentic and interpretation-free verification of these authors' results.

Rohde and Muller based their approach on FSA, whereas I focus on Gauss-Vaníček spectral analysis (GVSA). Besides geology and paleontology, virtually all science and engineering disciplines could benefit from the approach I describe. The main advantages of GVSA are in period detection from gapped records and in straightforward testing of the statistical null hypothesis. Unlike FSA, which draws periods from complete records of equispaced data, GVSA models the periodicity (via least-squares fitting the data with trigonometric functions) in raw and gapped records.

## Altered Records

Rohde and Muller reported that they used a stepwise time series $l_i \in \Re$ of finite size $n$ and a 50,000 years (50 Kyr) sampling interval for their data set.[1] (In fact, the sampling interval they used to come up with Figure 1 turned out to be 250 Kyr.) Their input record $l_i = l_j + l_k$ consisted only of some locally constant subsamples (100 Kyr to 500 Kyr), such that $l_{j+1} - l_j = 0$, and of some locally non-constant subsamples, such that $l_{k+1} - l_k \neq 0$, where $i = 1 \ldots n$; $j = 1 \ldots m$; $k = 1 \ldots p$; $n \in \aleph, m \in \aleph, p \in \aleph, m + p =$

MENSUR OMERBASHICH
*Berkeley National Laboratory*

*n*. The latter type of subsamples indicated *locally* that a genus variation had occurred.

A closer look at the data Rohde and Muller spectrally analyzed to get to Figure 1 reveals that the ratio of locally constant versus locally non-constant samples is more than 27:1. Such a situation arises as a direct consequence of using FSA, which requires that the data be equispaced and the number of data values be equal to a power of two. Rohde and Muller altered the original record by replicating more than 90 percent of the variation data to complete the inherently incomplete record over its entire span. The record these authors constructed was thus unavoidably characterized by false self-consistency of genera because all the genera were considered. Such a record depicts locally the genus invariability more than it depicts genus variation—in fact, genera variations should characterize the processed record, all the way through. This made the need to "prepare" (distort, actually!) data to be fit for FSA, as well as the need to boost the FSA spectral power in cases like this one, very obvious.

Rohde and Muller altered the record they had "completed" in this manner once again—this time by padding the already altered Sepkoski time series with zeroes—to make the record fit a required length. This also helps boost the FSA spectral power, which artificially completed records can affect in unpredictable ways. Rohde and Muller then manipulated their original record for a third time—by subtracting the arbitrarily selected (third-order) polynomial—which caused the record to largely retain its imposed self-consistency.

Clearly, any spectral analysis of such a manipulated record is subjective: altering data in three unrelated ways prevents an optimal estimate, and such approaches shouldn't be favored over those that analyze raw data directly. Besides, FSA offers no guarantee that detrending by a polynomial has a (physical) meaning in any realm outside FSA's. Rohde and Muller's intention to abridge FSA's shortfalls[3] by thrice altering the original data was neither entirely justifiable nor unbiased. In fact, as we'll see, such an approach can often be detrimental to results obtained in an FSA of gapped records containing natural data.

### Gauss-Vaníček Spectral Analysis

The preparation of a time series for FSA, which is currently the most widely used spectral analysis technique in all of the sciences, requires certain data manipulations—for example, researchers must artificially equispace (innately) gapped records, pad such records with zeroes to improve peak definition and peak resolution, and so on.

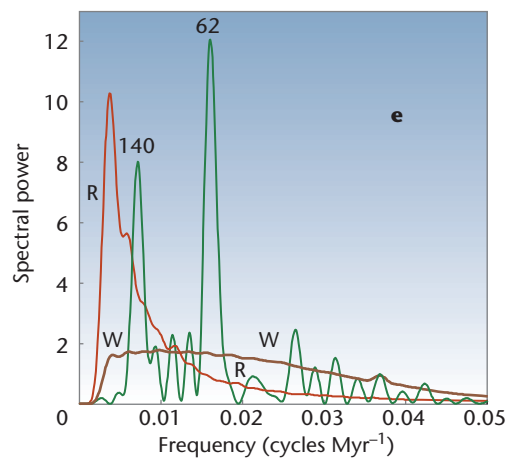

Figure 1. Time series sequence. Figure reprinted from R.A. Rohde and R.A. Muller, "Cycles in Fossil Diversity," *Nature*, vol. 434, 10 Mar. 2005, pp. 208–210.

Fortunately, other existing methods can produce estimates of the periodicities from raw, non-equispaced (discontinuous) records more accurately than FSA can.[3]

GVSA,[4,5] for example, can fit any (equispaced or not) time series with trigonometric functions and provide rigorous testing of the statistical null hypothesis.[3] It can also handle gapped records of any length, describe fields uniquely and relatively because of its output's linear background, and produce spectra in percentage variance (var%) or decibels (dB). Furthermore, GVSA's ability to remove systematic noise from a time series with minimal distortion of the spectrum of the remaining series has also been demonstrated.[6] These properties are particularly important for long records of natural data because, unlike GVSA, FSA generally boosts long-periodic noise in long gapped records.[3] (Here, long records are those that span one or more orders of magnitude longer in their time intervals than the periodicity claimed.[7]) GVSA is of particular use for the physical sciences as a field descriptor because it offers accurate simultaneous detection of field eigenfrequencies and relative dynamics.[8]

Researchers have used GVSA for the past 30 years in various fields, including astronomy, medicine, finance, geophysics, and mathematics, and as such, it's referred to in various ways: as Vaníček spectral analysis,[6] least-squares spectral analysis (LSSA),[7,8] and Lomb-Scargle or just the Lomb spectral analysis.[9] The latter name for the same method refers to GVSA simplifications

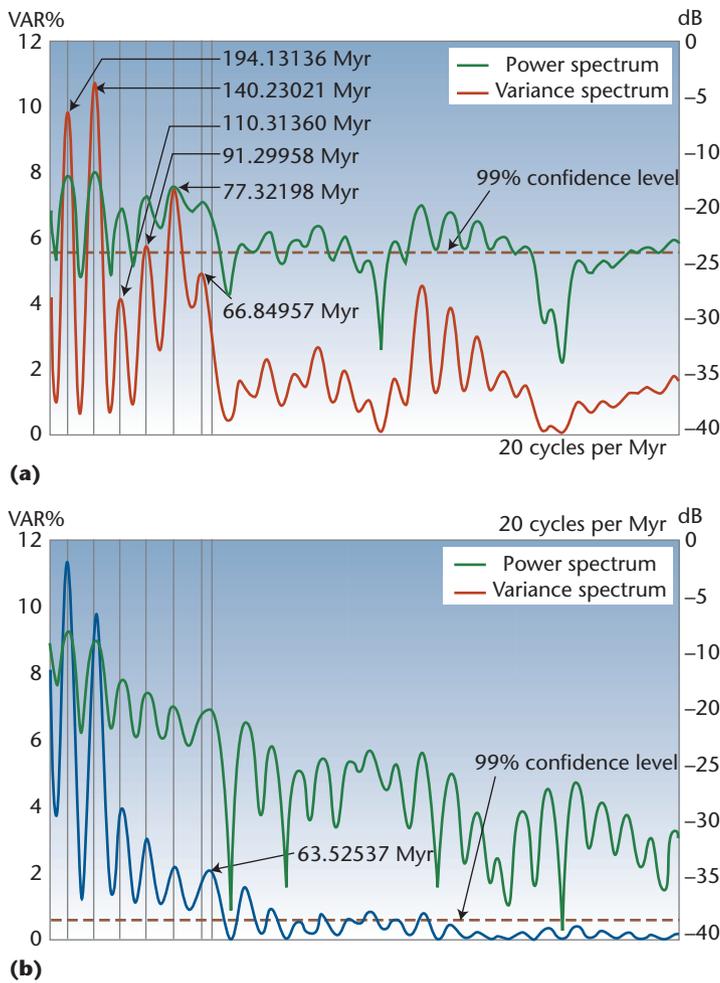

**Figure 2.** Gauss-Vaníček variance spectrum (blue) and power spectrum (green) of the Sepkoski compendium.[2] I performed Gauss-Vaníček spectral analysis (GVSA) of (a) well-resolved genera and (b) well-resolved genera distorted by Robert A. Rohde and Richard A. Muller's data manipulations.[1] Note the difference between the power spectra: in (b), the approach Rohde and Muller used boosts spectral power to a nearly 100 percent increase in signal decibel range, thus introducing at least 13 new falsely significant periods.

"devised" years after Vaníček originally developed GVSA.[4,5]

GVSA's attractiveness lies in its ease of use: its application is undemanding, with little to no data preprocessing required. Because it's variance-based, the method provides a straightforward statistical analysis with a generally linear depiction of background noise levels, and the output spectra require no postprocessing whatsoever. GVSA enables accuracy generally unattainable with FSA because of the way it handles data. Moreover, it's more reliable because of its significantly complete modeling of noise; in contrast, FSA merely unveils periodicity in presumably fit and undistorted data. GVSA was, in the past, relatively unknown to the broader scientific community because of its somewhat slower computer execution, but given the kind of computing power available today, as well as remarkable benefits from using GVSA, this drawback is almost a nonissue.[8]

### Results

When I use GVSA to process the rigorously considered (noncontinuous, nonzero-padded, and not arbitrarily detrended) Sepkoski record—regarded here as the true record of genera variations—the results observed in Figure 2a and Tables 1 and 2 emerge.

Note that input data are regarded here as the fundamental criterion of a physical analysis result's validity. Adopting this as the universal principle of physics, no other comparison is needed, thus, by definition, GVSA is superior to FSA. Moreover, we don't need additional proofs other than independent verification of GVSA on synthetic data.[8] (By definition then, processing the data "as are" should represent the standard by which all other spectral analysis methods must be verified, especially when other methods require altering data to prepare data for processing or when other methods require manipulations of output to be able to perform any interpretation at all.)

When I remove the zeroes and replicated data from the original series Rohde and Muller altered,[1] GVSA of the restored Sepkoski record doesn't produce the claimed 62 ± 3 Myr period, but it does detect the 194 Myr and 140 Myr periods other researchers have reported.[1] A period found among all the noise that's closest to Rohde and Muller's result is 66.9 Myr, which falls well out of their claimed accuracy of ±3 Myr and under their claimed confidence level of 99 percent. For considerations that are even more stringent, we should also note the computed statistical fidelity, Φ. Anything below a threshold of Φ = 12.0 is generally noise. Only the 194 Myr and 140 Myr periods, here extracted from GVSA of the restored Sepkoski compendium, remain above that threshold (see Table 1).

Figure 2b shows that the period Rohde and Muller claim as real is part of the overall noise—at least 13 periods were artificially created by the boost,[2] all climbing abnormally to above the 99 percent confidence level (compare with Figure 1). When compared to Figure 2a, the approach used in Figure 2b boosts the spectral power up to a nearly 100 percent increase in the decibel signal range. Thus Rohde and Muller introduced the

| Table 1. Spectral peaks found in the gapped Sepkoski series restored by deleting all zeroes and repetitive data from Robert A. Rohde and Richard A. Muller's record.[1] | | | |
|---|---|---|---|
| Period (Myr) | Fidelity (Phi) | GVS (var%) | Power (decibel) |
| Above 99% confidence level at a 5.46 var% | | | |
| 194.13136 | 35.02508 | 9.8151 | –9.63239 |
| 140.23021 | 18.27557 | 10.6993 | –9.21501 |
| 91.29958 | 7.74685 | 5.7162 | –12.17329 |
| 77.32198 | 5.5564 | 7.4639 | –10.93347 |
| Below 99% confidence level at a 5.46 var% | | | |
| 110.31360 | 0 | 4.1056 | –13.68413 |
| 66.84957 | 0 | 4.8538 | –12.92313 |
| 32.13252 | 0 | 4.4945 | –13.27350 |

| Table 2. Spectral peaks found in Robert A. Rohde and Richard A. Muller's zero-padded data.[1] | | | |
|---|---|---|---|
| Period (Myr) | Fidelity (Phi) | GVS (var%) | Power (decibel) |
| Above 99% confidence level at a 0.42 var% | | | |
| 190.72165 | 33.57153 | 11.2937 | –8.95119 |
| 137.56541 | 17.46584 | 9.7092 | –9.68462 |
| 108.65782 | 10.89665 | 3.8862 | –13.93265 |
| 90.16245 | 7.50279 | 2.9691 | –15.14288 |
| 77.04766 | 5.47886 | 2.1547 | –16.57147 |
| 63.52537 | 3.72448 | 2.0130 | –16.87334 |
| 54.85997 | 2.77768 | 1.5219 | –18.10948 |
| 49.59786 | 2.27037 | 0.8694 | –20.56968 |
| 45.25686 | 1.89034 | 0.5019 | –22.97196 |
| 41.85520 | 1.61685 | 0.6761 | –21.67031 |
| 38.58335 | 1.37395 | 0.6015 | –22.18157 |
| 36.44655 | 1.22598 | 0.7738 | –21.07999 |
| 34.15385 | 1.07659 | 0.5662 | –22.44537 |

new, false, "99 percent significant" periods detected in the corresponding variance-spectra of noise.

## Discussion

If the period Rohde and Muller claimed were real, GVSA of the restored Sepkoski compendium would produce that period as well, along with the 194 Myr and 140 Myr periods. However, this wasn't the case, and it's the main reason for questioning the claimed period or any other periods discovered in the same way (that is, via arbitrarily detrended, several times distorted, data).

By definition and its limitations, FSA requires an unbiased sample of equispaced data to produce unbiased estimates of periodicity. This is what makes FSA an entirely inappropriate tool for drawing periods from the original Sepkoski record. By itself, the zero-padding procedure normally doesn't result in a bias in the estimates, but for zero padding to induce no bias, the signal has to be causal and the record fully populated by the data of exactly that category we want to study. However, Rohde and Muller never discuss causality of their data or of portions of data, nor do they use a record that is fully populated by data belonging to the same category as the sought-for periods.[1] The authors admittedly discuss biases in their data as a possible cause for the claimed 62 Myr period, but their results can't be considered impartial because they disregarded important properties of the way in which they treated the data.

They also don't report the difference in the reported period when the artificially equispaced record is processed with FSA instead of something like GVSA. When using GVSA to analyze the record they distorted, the period closest to the claimed period is 63.5 Myr as in Figure 2b (although Figure 2b and Figure 1 aren't entirely com-

JULY/AUGUST 2006                                                                                                                                                    29

parable, since their domains of application may not be the same). Two different analysis techniques applied on the same data set must produce the same result, within the realms of the respective techniques' applicability. The difference between results must be, at best, less than the significant digit to which the claim is laid; at worst, it must be less than the claim's labeled uncertainty. This constitutes a technique-independent verification of a reported scientific result.

Comparison between Figure 2a and Figure 1 indicates a failed verification at the claimed 99 percent confidence level. In addition, the approach Rohde and Muller took when assessing their claimed period's statistical significance is problematic: specifically, the assumption they applied—that the diversity changes reflect only random walk—contradicts the claim of the period itself and to it associated the abrupt changes in diversity every 62 Myr. It appears that they applied an induction-proof logic uncritically (directly) from mathematics onto a physical science problem.

Unlike FSA, GVSA made no distortions of the raw data, used no assumptions such as that on the quality of trend (removal), and added no arbitrary data to either augment the data so that they exhibit unexpected periodicities or that the spectral technique itself performs unnaturally well.

The failed verification described here illustrates the importance of choosing numerical analysis techniques that are most appropriate for processing the data of interest. In short, any scientific or engineering endeavor aimed at learning about periodicity in data could benefit from the GVSA method. Its chief advantage is its ability to process time series that aren't uniformly spaced in time, such as most natural records. Besides analyzing incomplete records, researchers might also want to remove less trustworthy data from any time series before analyzing it with the Gauss-Vaníček method.[10] This could increase both the accuracy and reliability of spectral analyses in general. Other important advantages of GVSA include straightforward testing of null hypothesis, as well as accurate and simultaneous detection of eigenfrequencies and relative dynamics in physical sciences.[11]

## Acknowledgments

*I thank the three anonymous reviewers and Spiros Pagiatakis (University of Toronto) for his software LSSA v.5.*

***Mensur Omerbashich** is an assistant professor of geophysics in the department of physics at the University of Sarajevo. His research interests include the ramifications of theoretical geophysics for theoretical physics, physics at Planck's versus Newtonian scales, and microgeodesy/nanogeodesy. Omerbashich has a PhD in theoretical geophysics from the University of New Brunswick. He is a member of the American Geophysical Union. Contact him at momerbasic@pmf.unsa.ba, cc: omerbashich@ yahoo.com.*

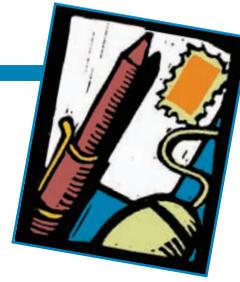

# THE FIRST AND LAST WORDS

Francis Sullivan's thoughts on programming errors ("Wrong Again!" vol. 9, no. 3, p. 96) speculate that formal verification of computer programs is unlikely. It's worth noting that Bertrand Meyer's theory of "design by contract" was inspired by the work of theorists in formal proofs of programs, but being a person in need of something that actually helped him get programs right, rather than producing papers in journals, he came up with a very workable approach. DBC consists of a language for assertions that can be optionally checked at runtime, backed with a theory about DBC's relationship to inheritance and user–supplier relationships. In his language (Eiffel), Meyer was able to make this approach fully integrated into the language. In other languages, only a poor imitation can be achieved, and yet that poor imitation is often very powerful. I have found it by far the most effective approach to correctness for scientific computing.

When I helped write the Eiffel mathematics library, EiffelMath, which encapsulated the NAG C library for use in Eiffel, I was amazed at the effectiveness of the approach. I subsequently used the ideas in C++ projects to great effect. My understanding is that this was planned for Java but omitted due to a rush to vend. If so, that's a real pity. I urge computational scientists to familiarize themselves with this theory.

I also read your First Word ("You're Recommending What?!" vol. 9, no. 3, p. 2), and got a big kick out of it. Today at work, we were discussing the younger members of the team who've decided that they don't want or need a developer's manual but instead want a Wiki. Apparently they can't read off paper any more.

An Internet bridge partner of mine is married to a famous Canadian politician. She's named Julia and has two children, but her Wikipedia entry said she was Joan and had three. I wrote her an email and said maybe there was something going on that she didn't know about; when we all finished laughing, of course, her son knew how to fix it.

I was jokingly going to tell you that the next thing that would happen is videos instead of user manuals, but it already happened. For example, http://showmedo.com/videos/python has more than 90 videos about Python, including one on Django. Some are in German, but that's OK—the kids just look at the pictures.

Time to retire, because I don't get it.

*Paul F. Dubois*
*paul@pfdubois.com*

## A Plea for Python

I'm sure I'm not the only one who finds it amusing that in an entire issue devoted to the benefits of Python for scientific computing, the magazine offers a book review of a text that uses four different computer languages, none of which is Python. I'd feel a lot better about adopting a new course in my computing if I could find a good book on scientific computing with Python and its tools and libraries. I've found a few tutorials but nothing substantial. Are any authors rushing such a book into production? How many college campuses are using this for their course work?

When you next visit the Python topic, please list numerical books and college courses in which Python is the basis for performing the computations.

*Robert Love*
*rblove@airmail.net*

## Gauss-Vaníček or Fourier Transform?

From J. John Sepkoski's record of the marine animal genera that appear in the overall fossil record,[1] Robert Rohde and Richard Muller[2] extracted a time series of marine animal diversity and analyzed it using the discrete Fourier transform. They found a highly significant 62-million-year periodicity in the series, detrended by a cubic. This time series has 167 data points, and the times aren't uniformly distributed, so to use the Fourier transform, Rohde and Muller assumed marine diversity to be constant between data points and evaluated it at times equally spaced at 0.25 million years to obtain a series of some 2,170 terms. They then fit a cubic polynomial to that series and examined the residuals—the series minus the cubic. Finally, they extended the residual series with zeroes so that the extended series' Fourier power spectrum would have densely distributed frequencies and to enable the use of the

**Send letters to**

Jenny Stout, Lead Editor
jstout@computer.org

Letters are edited for length, clarity, and grammar.

fast Fourier transform. Thus, the original series of 167 terms appears as a ghost in the series for which the Fourier power spectrum (FPS) was actually computed.

The Gauss-Vaníček power spectrum (GVPS) of a time series is a measure of how well various frequencies' harmonic functions fit the time series, doesn't require data to be equally spaced in time, can be computed for every frequency without needing to extend the original series, and is possibly more suitable than the Fourier transform for analyzing time series such as the marine diversity series. In an article that appeared in this magazine last year, Mensur Omerbashich[3] computed the GVPS of the nondetrended marine genera diversity time series and found that no significant 62-million-year periodicity appeared in the unmodified series, reporting instead significant periodicities of 194 and 140 million years.

But in fact, as I show on p. 61 of this issue, the Gauss-Vaníček spectral analysis of the diversity time series detrended by a cubic *exactly* matches Rohde and Muller's Fourier transform analysis. Moreover, neither the FPS nor the GVPS identifies any significant periodicity in the unmodified diversity time series. Thus, in analyzing the diversity time series, paleontologists need decide only whether to detrend the data: Gauss-Vaníček and Fourier spectral analyses yield identical results.

*James L. Cornette*
*cornette@iastate.edu*

### References

1. J.J. Sepkoski, "A Compendium on Fossil Marine Animal Genera," *Bulletin of Am. Paleontology*, no. 363, Paleontological Research Inst., 2002.
2. R.A. Rohde and R.A. Muller, "Cycles in Fossil Diversity," *Nature*, vol. 434, 10 Mar. 2005, pp. 208–210.
3. M. Omerbashich, "Gauss-Vaníček Spectral Analysis of Sepkoski Compendium: No New Life Cycles," *Computing in Science & Eng.*, vol. 8, no. 4, 2006, pp. 26–30.

*Mensur Omerbashich replies:*
Based on his own power-spectra (PS) of Sepkoski data detrended by a specific function, James Cornette imputes that detrending in general is the modification vital for assessing my paper and the FSA, and speaks as if I used Gauss-Vaníček power-spectra (GVPS) to draw my periods. Neither applies. First, the detrending is excluded from the Gauss-Vaníček spectral analysis (GVSA) when variance-spectra (VS) are used for verification purposes because detrending doesn't comply with the raw data-only physical criterion accompanying the GVSA by definition, and because variance feeds on noise. Noise is the most natural gauge available for this sort of verification, so more noise means clearer VS-PS separation; conventional noise thus becomes part of the signal for verification purposes. Second, the variance gives the most natural description of noise, whereas only VS can measure the signal imprint strength in noise uniquely, so it is GVVS, not GVPS, that are useful for SA method verification purposes. Then to draw periods, I used VS only, with their depicted 99 percent confidence levels. Cornette didn't ask for clarification, although my final printed article had the unsolvable color coding, with blue, red, and brown depicting the GVVS of Rohde-Muller-adapted data, the GVVS of unmodified (depadded and derepeated) and nondetrended data, and the GVVS's 99 percent confidence levels, respectively. I plotted all three entities in black in my original submission, with GVPS in gray. [*See the original illustration reprinted at the end of this letter.—eds.*] Demonstrably confused, Cornette then arbitrarily chose to compare GVPS with FPS on the unmodified and modified cubic-detrended data, respectively, equating such an unnatural (indiscriminative in manipulation type) mix of data treatment approaches with my comparison of the Rohde-Muller FPS of manipulated (modified and detrended) data versus the GVVS of non-manipulated (raw) data. Comparing FPS versus GVPS on the altered data only doesn't constitute a physically independent verification because this merely computes two PS that by definition (regardless of SA method) must react in the same way to the same data alteration (the detrending by a cubic), which they do, as Cornette trivially shows. Note I was denied preprint access to the result he states is "shown on p. 61," so my response could be incomplete in addressing his claim's technical aspects—the meaning of his "*exactly*" and VS if any.

The objection Cornette attempted to make opens topics that are more fundamental than the issue of whether he understood the GVSA and my paper entirely. Besides making a negative verification of a previous report claiming a new period in a fossil record, my paper showed that the paleontological paradigm, which *assumes* diversities are completely known (real and static once they occur), is unsound. This frail assumption is what allows researchers to fill sparse records by repeating (here, around 90 percent of) the data until the record is made fully "populated" and thus Fourier-ready. I challenged this paradigm by showing that PS of a manipulated data set can differ significantly from VS of the respective raw data. Cornette, on the other hand, computes FPS and GVPS to the best of the two methods' mathematical abilities (by twice using the cubic-detrended data) but not their natural abilities (by using the manipulated versus raw data). Based on those two PS giving statistically indistinguishable results while differing significantly from my result, he concludes that all one has to decide prior to selecting a SA method is whether one wants to indiscriminatingly detrend one's data or not. However, there is no reason why PS of modified should be the same as PS of

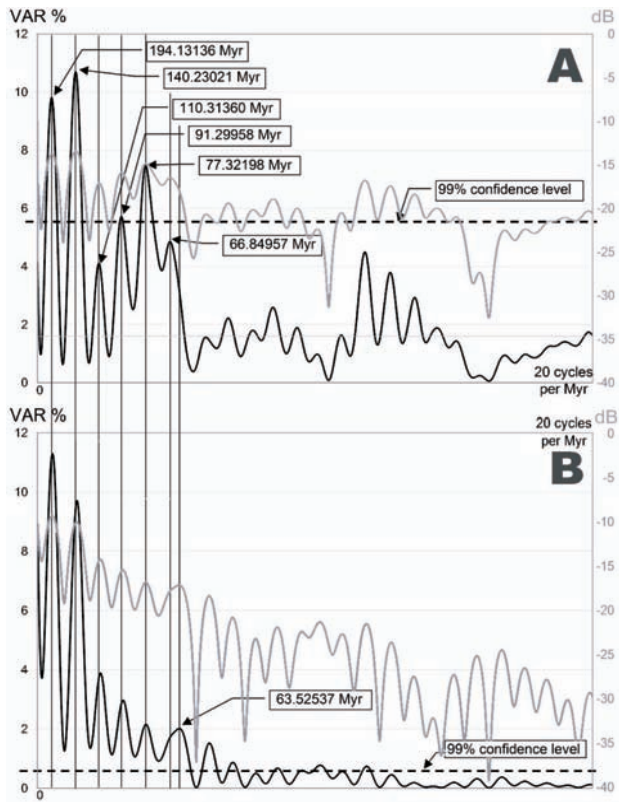

**Errata.** As Mensur Omerbashich mentioned, his original figure was accidentally discolored in the print version. We're reprinting his figure here, as he originally submitted it. See his reply for the original figure description. We apologize to the author as well as to the readers for any confusion this might have caused. —eds.

the respective unmodified non-raw data that were arbitrarily (say, other-than-cubic-) detrended; this would require that the detrending be universal, that is, insensitive to the type of a detrender function, which is a nonsense. Only if data repetition in inherently sparse paleontological records were tolerably natural (and not just another mathematical trick) would Cornette's "any detrending" concept make sense from the physical viewpoint, that is, beyond any doubt and in all real-life situations regardless of SA method. If that were the case here—if an unspecified combination of data repetitions and detrending weren't significantly affecting the spectrum—then the claimed period would have remained significant in VS of the non-manipulated Sepkoski record as well, just as the record's longest 194- and 140-million year-long periods have. It's chiefly because this wasn't the case that the claimed period doesn't seem real, and not just because its VS estimate differs a few percent from the claimed value. Despite this obvious inconsistency and based on his ill logics (it's unclear how he computes "FPS… in unmodified… time series" when the FSA can't even process gapped records), Cornette proposes we forge the entire approach to SA verification: "compare" the two methods by comparing FPS of somehow-completed and somehow-detrended versus GVPS of unmodified and somehow-detrended data. But two wrongs never add up to one right so *somehow* and *detrending* become the issue, and all his procedure does is verifying not the accuracy of the two methods themselves, but whether GVPS and FPS give precisely the same result when compared using the same vector of real numbers twice. Cornette thus objects to a fact: after applying an alternative method to the fullest of its abilities, I primarily challenged a disputable (method-driven) paradigm of a posteriori static diversities. I didn't examine the effects of the detrending as stand-alone but as applied on an extremely sparse record—a situation likely to contribute significant noise. It's a combination of the detrending and data sparseness that could have had influenced the claimed result significantly; Cornette had no alternative ways of checking for such combined effects because VS represent the only means to do that, and he diverged to using PS only.

Because paleontology relies on the paradigm that I challenged, at long last it might be obvious why climatology is so dreadfully controversial: perhaps we've hit the wall when it comes to the reliability of the tools used for analyzing sparse (erratic?) records. Any manipulations we perform on raw data actually manipulate the public as a whole. Still most researchers pretend they know what's going on when treating sparse or otherwise hard-to-understand records (for instance, what's the cost of "detrending," "ghosts," and other poorly understood procedures and phraseology in the FSA?). Instead of trying to understand (the limits on usefulness of) such data, we began at some point with "modeling" (arbitrarily adjusting, in fact) our own understanding of such data, which we then sell to the public as the analyses of the data themselves; no wonder controversies abound. Mathematicians like Fourier (Cornette) often create imaginary worlds in which, quite Nietzscheanly, the cause justifies the manipulation. Such concepts ambush all physical sciences ever since A. Einstein half-jokingly proposed overthrowing H.R.M. king Data. But just as its indefinability prevents the fake king Time from being re(oy)ally spasmatic, so are only the raw data—thy king.